\documentclass{PoS}
\title{Stellar populations and star formation in AGN hosts at intermediate redshift in the SHARDS survey}
\ShortTitle{Stellar populations and star formation in AGN hosts at intermediate $z$}
\author{\speaker{Antonio Hern\'an-Caballero}, Almudena Alonso-Herrero\\
        Instituto de F\'isica de Cantabria, Spain\\
        E-mail: \email{ahernan@ifca.unican.es}, \email{aalonso@ifca.unican.es}}

\author{Pablo G. P\'erez-Gonz\'alez, Antonio Cava, Nicol\'as Cardiel\\    
        Universidad Complutense de Madrid, Spain\\      
        E-mail: \email{pgperez@fis.ucm.es}, \email{acava@fis.ucm.es}, \email{cardiel@fis.ucm.es}}
        
\author{and the SHARDS team}

\abstract{SHARDS is an ongoing ESO/GTC large program that is obtaining ultra-deep photometry of the GOODS-North field in 24 medium-band filters (reaching m=26.5 AB in all bands) in the 500--950 nm range with GTC/OSIRIS. It is designed to study the properties of high-$z$ massive galaxies, but it can also provide very valuable information about the population of AGN at intermediate redshifts ($z$$\sim$0.5--2). Here we present preliminary results on a study of the stellar populations and star formation activity in the host galaxies of X-ray selected optically faint AGN at 0.6$<z<$1.1. We demonstrate that the SHARDS photometry provides a reliable measurement of the 4000 \AA{} break (Dn(4000), an indicator for the age of stellar populations) down to m=26.5 AB. We confirm that most X-ray selected AGN are hosted by massive galaxies (typically M$_*>$10$^{10.5}$ M$_\odot$) and that the fraction of galaxies hosting an AGN increases with the stellar mass.
AGN-hosts have restframe U-V and Dn(4000) comparable to those of non-active galaxies of the same mass, that is, they do not appear to have on average younger stellar populations, unlike in the Local Universe.
Nevertheless, $z$$\sim$1 AGN hosts show an excess of IR emission at $\lambda$$>$3$\mu$m compared to non-AGN galaxies which might indicate increased star formation rates. In addition, the frequency of AGN in massive galaxies is about twice higher for the ones with young stellar populations compared to the older ones.}

\FullConference{Nuclei of Seyfert galaxies and QSOs - Central engine \& conditions of star formation\\

                 November 6-8, 2012\\

                 Max-Planck-Insitut f\"ur Radioastronomie (MPIfR), Bonn, Germany}

\begin{document}

\section{Sample selection}
\vspace{-0.1cm}
We obtained redshifts, restframe colors and stellar masses for the SHARDS sources \cite{Perez-Gonzalez13} by cross-identifying to the GOODS-N catalog from \cite{Perez-Gonzalez08}.
We selected the 1667 SHARDS sources with 10$^{9}$$\leq$M$_*$/M$_\odot$$\leq$10$^{12}$ and
0.65$<$$z$$<$1.07. 
From these, 57 are included in the Chandra 2 Ms catalog from \cite{Alexander03}.
To compare the AGN hosts with the general galaxy population, we focus on the 53 \mbox{X-ray} sources with a clear 1.6$\mu m$ bump. That is, the AGN does not dominate the near-IR emission and we can study their host galaxies. These are thus moderate luminosity AGN (see also \cite{Alonso-Herrero08}).
\vspace{-0.1cm}
\section{Results}
\vspace{-0.1cm}
We measure the 4000 \AA{} break following \cite{Balogh99}.
Due to the lower resolution of the SHARDS SEDs (R$\sim$50) compared to spectroscopic samples, a calibration with spectra is required. The total uncertainty in Dn(4000) after calibration is $\lesssim$10\% for most sources.
 We find that most X-ray selected AGN are hosted by massive galaxies (typically M$_*>$10$^{10.5}$M$_\odot$, see also \cite{Alonso-Herrero08}), and that the fraction of galaxies hosting an AGN increases steeply with the stellar mass, from $<$5\% at M$_*$$<$10$^{10}$M$_\odot$ to $\sim$30\% at M$_*$=10$^{11.5}$M$_\odot$ (Fig. 1, top left).
For normal galaxies, both \mbox{U-V} and Dn(4000) correlate with M$_*$ (Fig. 1, right), indicating that more massive galaxies have older stellar populations. The \mbox{U-V} color and Dn(4000) of AGN-hosts are comparable to those of normal galaxies with the same mass, and place them in the green valley and red sequence. AGN also occupy the same locus as normal massive galaxies in the \mbox{U-V} vs Dn(4000) diagram (Fig. 2, left), that is, they do not appear to have on average younger stellar populations than non-active galaxies, unlike in the Local Universe \cite{Kauffmann03}.
The distribution of Dn(4000) values for AGN and non-AGN galaxies are very similar once the mass dependence of the AGN frequency is taken into account (Fig. 1, left bottom), and a KS-test indicates no statistically significant differences in the stellar populations of both subsamples. Nevertheless, for massive (M$_*>$10$^{10.5}$M$_\odot$) galaxies the probability of hosting and AGN decreases with increasing average stellar age, and is about twice as high for the younger (Dn(4000)$<$1.2) galaxies compared to older (Dn(4000)$>$1.6) ones. In addition, AGN hosts show an excess of IR emission at $\lambda$$>$3$\mu$m with respect to non-AGN galaxies of the same mass which might be interpreted as an excess of current SFR (Fig. 2, right), albeit some degree of contamination from emission of the AGN torus is also likely.

\begin{figure}
\begin{center}
\includegraphics[height=7.0cm,width=5cm]{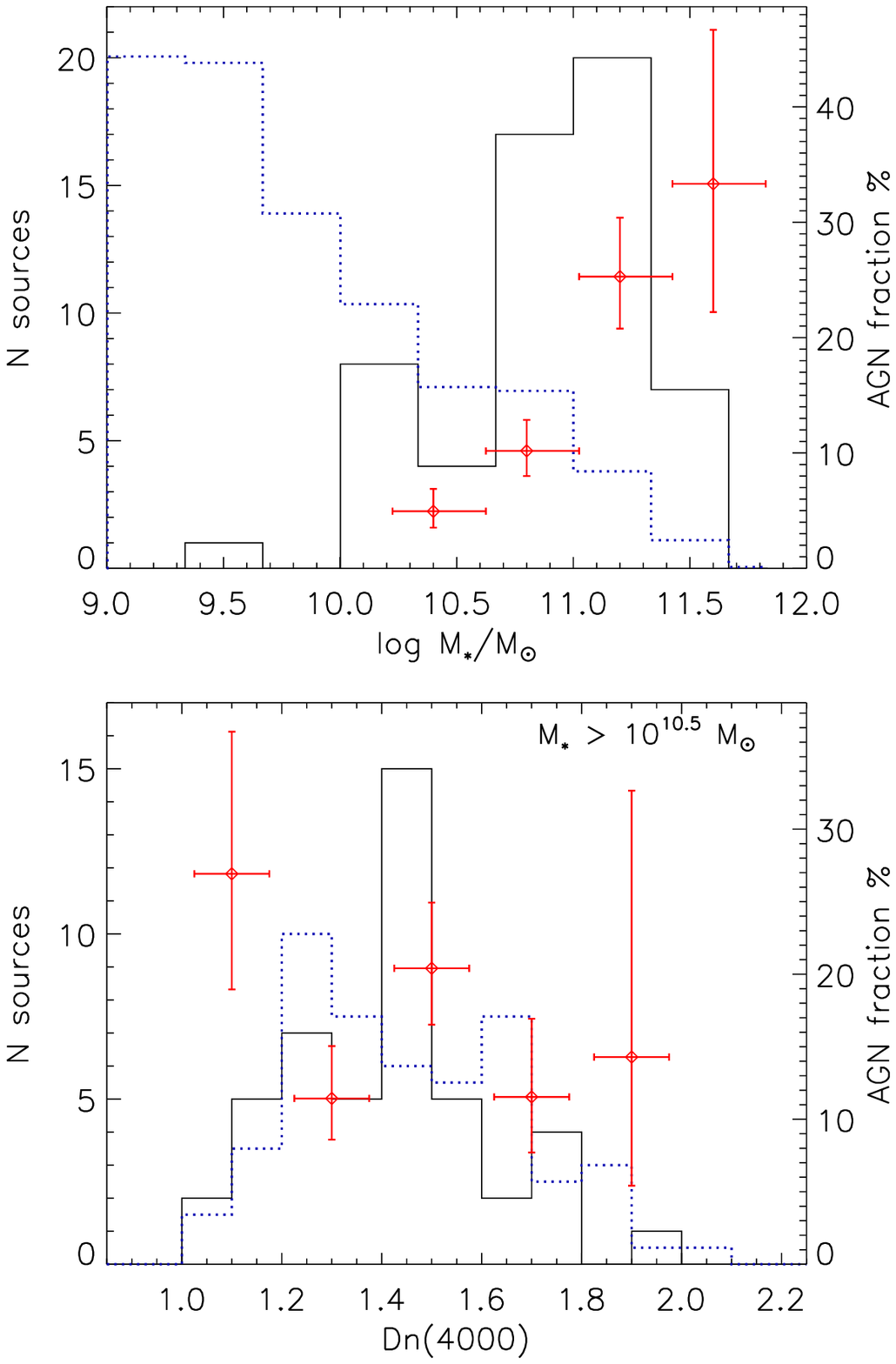}
\hfill
\includegraphics[height=7.0cm,width=9.5cm]{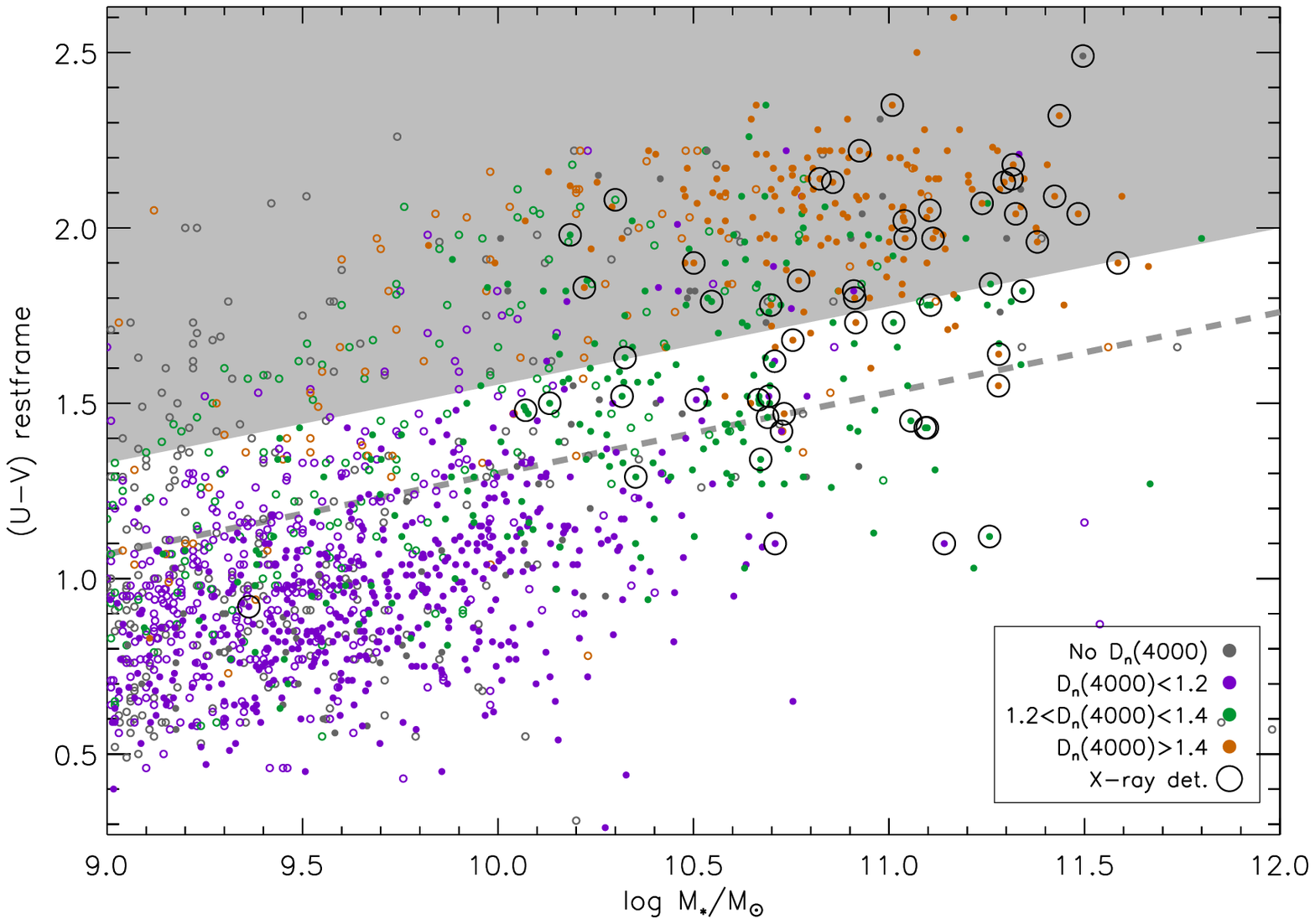}
\end{center}
\caption[]{(Top left): stellar mass distribution of the AGN hosts (solid line) and all galaxies (dotted line, scaled by a factor 0.05). Solid symbols with error bars indicate the AGN fraction as a function of M$_*$. (Bottom left): distribution of Dn(4000) values for AGN hosts (solid line) and a random sample of non-AGN galaxies with the same mass distribution (dotted line). Solid symbols with error bars indicate the AGN fraction among M$_*>$10$^{10.5}$M$_\odot$ galaxies as a function of Dn(4000). (Right): U-V vs stellar mass with color coding for the Dn(4000) index. Solid (open) symbols represent sources with spectroscopic (photometric) redshift. Big circles enclose AGN host galaxies. The shaded area represents the red sequence above the cut defined by \cite{Cardamone10} for 0.8$<$$z$$<$1.2, while the dashed line indicates the approximated location of the green valley.\label{f34}}
\end{figure}

\begin{figure}
\begin{center}
\includegraphics[width=7.5cm,height=6.0cm]{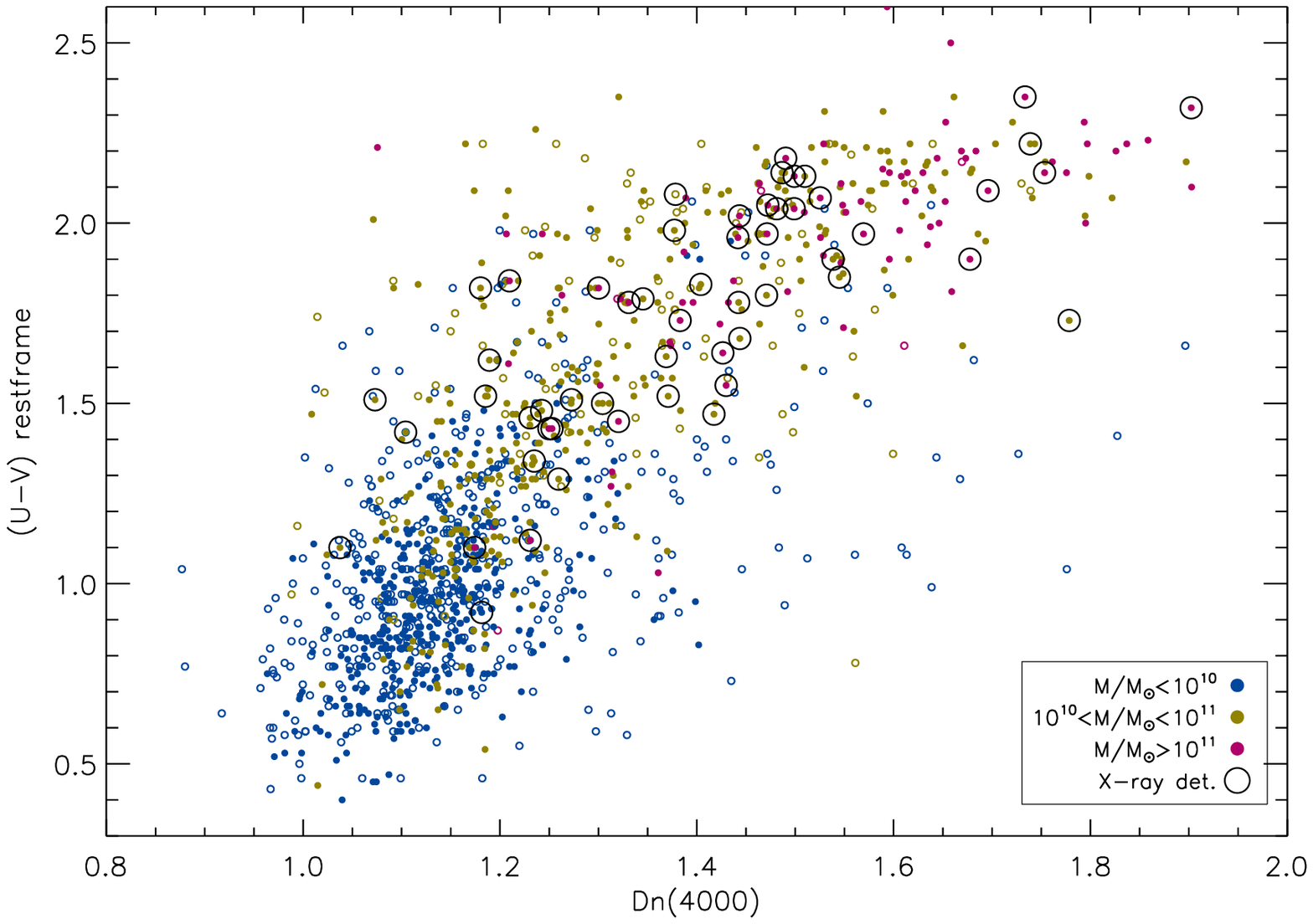}
\hfill
\includegraphics[width=7.5cm,height=6.0cm]{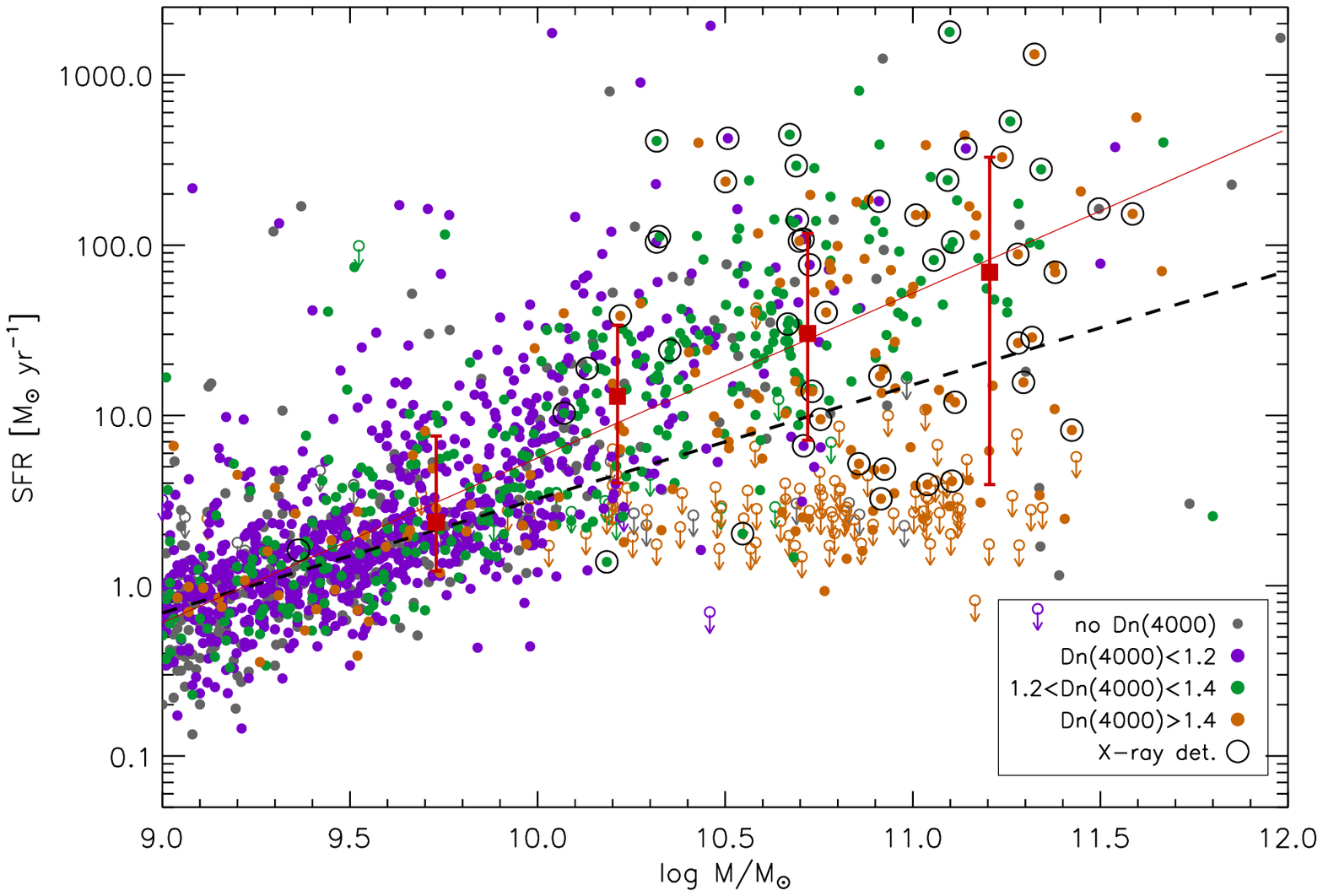}
\end{center}
\caption[]{(Left): U-V vs Dn(4000) with color coding for stellar mass. Symbols as in Fig. 1. (Right): SFR estimate from the UV+IR emission vs stellar mass with color coding for the Dn(4000) index. Sources detected (undetected) at 24$\mu m$ are shown with solid (open) symbols. Big squares represent average values for sources with SFR estimates (no upper limits) in bins 0.5 dex wide and the solid line is the best linear fit of $\log$(SFR) vs $\log$(M$_*$). For reference, the `main sequence' from \cite{Noeske07} is shown in dashed line.\label{UV-plots}}
\end{figure}
\noindent\textbf{Acknowledgement} AHC and AAH acknowledge support from the Augusto G. Linares program.

\end{document}